# Ultrathin film of 3D topological insulators by vapor-phase epitaxy: Surface dominant transport in wide temperature revealed by Seebeck measurement

Stephane Yu Matsushita,[1*] Khuong Kim Huynh,[2] and Katsumi Tanigaki[1,2 **]

[1]*Department of Physics, Graduate School of science, Tohoku University, Sendai 980-8578, Japan*

[2]*WPI-Advanced Institute for Materials Research, 2-1-1 Katahira, Aoba-ku, Sendai, Miyagi, 980-8578, Japan*

*E-mail address: *m.stephane@m.tohoku.ac.jp, **tanigaki@m.tohoku.ac.jp

PACS numbers: 73.25.+i, 73.50.Lw, 72.20.My

## Abstract

Realization of intrinsic surface dominant transport in a wide temperature region for a topological insulators (TIs) is an important frontier research in order to promote the progresses of TIs towards the future electronics. We report here systematic measurements of longitudinal electrical transport, Shubnikov-de-Haas (SdH) quantum oscillations, Hall coefficient ($R_\text{H}^\text{2D}$), and Seebeck coefficient as a function of film thickness ($d$) and temperature using high quality Bi$_{2-x}$Sb$_x$Te$_{3-y}$Se$_y$ (BSTS) single crystal thin films grown by physical vapor-phase deposition. The thickness dependence of sheet conductance and Seebeck coefficient clearly show the suppression of semiconducting hole carriers of bulk states by reducing film thickness, reaching to the surface dominant transport at below $d_\text{c}$=14 nm. Quantitative arguments are made as to how the contribution of itinerant carrier number ($n$) can be suppressed, using both $R_\text{H}^\text{2D}$ ($n_\text{Hall}^\text{2D}$) and SdH ($n_\text{SdH}$). Intriguingly, the value of $n_\text{Hall}^\text{2D}$ approaches to be twice of $n_\text{SdH}$ below $d_\text{c}$. While $R_\text{H}^\text{2D}$ shows a negative sign in whole temperature region, a change from negative to positive polarity is clearly observed for $S$ at high temperatures when $d$ is thick. We point out that this inconsistency observed between $R_\text{H}^\text{2D}$ and $S$ is intrinsic in 3D-TIs and its origin is the large difference in carrier mobility between the bulk and the topological surface. We propose that Seebeck coefficient can become a convenient and powerful tool to evaluate the intrinsic carrier concentration for the topological surface in 3D-TIs even in the absence of magnetic field.

# 1. Introduction

Topological insulators (TIs) have currently been attracting much attention from the viewpoint of contemporary materials science generating new electronic states, such as gapless helical massless Dirac fermions on two-dimensional (2D) surface or one-dimensional (1D) edge [1-3]. The existence of such special energetic states on the topological surface states has unambiguously been confirmed by surface sensitive measurements of angle and spin resolved photoemission spectroscopy [4-7]. Although many theoretical approaches suggest exotic physical properties as well as novel applications of TIs, clear clarification of such physical properties has still been difficult experimentally because itinerant carriers thermally generated from the bulk bands are frequently involved in experimental observations of physical properties. Therefore, one of the most important requirements in order to unveil the intrinsic physical properties of TIs is how we can evaluate the physical properties by minimizing and discriminating the contribution of bulk carriers when we measure the properties of topological surface Dirac states (TSDS). This can be realized in principle by either tuning the Fermi level ($E_F$) inside the bulk gap or growing high quality ultra-thin films to reduce the bulk contribution in total. The $E_F$ is known to be engineered in synthesis by the concept of charged defects controlling, and two kinds of highly bulk insulating 3D-TIs of $Bi_{2-x}Sb_xTe_{3-y}Se_y$ (BSTS) and $Sn-Bi_{1.1}Sb_{0.9}Te_2S$ (Sn-BSTS) are presently proposed [8-14]. As for the reduction in film thickness, on the other hand, the suppression of the bulk in thin films can be recognized only at low temperatures, but no systematic study has been carried out and common experimental consensus has not yet been made to confirm whether surface dominant transport can be realized in a wide range of temperature.

In general, Hall measurement is a common and useful technique to evaluate contributions of conductive channels in electrical transport. Whether electrical conduction of a material is made via either of single or multi carriers could accurately be judged by evaluating the transverse electrical transport $R_{yx}$ ($B$) (Hall effect: $R_H$): a linear progression as a function of $B$ is essential for single-channeled carriers, while a non-linear one is observed for multi-channeled carriers. The non-linear term of $R_{yx}$ ($B$) of 3D-TIs, caused by not only the large mobility difference between the surface and bulk carriers but also the robust topological protection, can only be evident under high $B$ above 10 T. It is not experimentally easy for interpreting these experimental data to deduce a firm conclusion as to whether predominant properties resulting from TSDS are observed from the linear dependence of $R_{yx}$ ($B$) and therefore debate still continues. The surface dominant electrical transport can qualitatively be discussed from the temperature ($T$) dependence of longitudinal and transverse electrical transport or the thickness ($d$) dependence of sheet resistance ($R_\square$). In principle, more accurate analytical discussion can be made by employing both non-linearity in $R_H$ and Shubnikov-de-Haas (SdH) quantum oscillations under extremely high $B$ field [15].

However, valid combined-measurements of $R_H$ and SdH are required to be carried out at low $T$ below 10 K under high $B$, and therefore do not allow one to make direct discussions at room temperature.

One another method to distinguish the carrier type is to measure the Seebeck coefficient ($S$) [16]. The polarity of $S$ reflects the polarity of the dominant carrier as well as $R_H$, i.e. positive for p-type carrier and negative for n-type carrier. Moreover, it is able to distinguish whether the carrier is semiconducting or metallic by measuring the temperature dependence of $S$: $S$ shows non-linear $T$ dependence for semiconducting carrier, while it shows leaner $T$ dependence for metallic carrier. Recently, we have observed the thickness dependence of Seebeck coefficient in BSTS films and revealed the surface dominant $S$, which can be judged from the different polarity and temperature dependence of p-type semiconducting carriers and n-type metallic carriers. The former originates from the bulk and the latter from the topological surface [17]. Measurements of S in 3D-TIs thin films have also been carried out in BiSbTe (BST) alloy. J. Zhang *et al.*, tuned $E_F$ by chemical doping on BST thin films and observed an inconsistency in polarity observed between the Hall and the Seebeck coefficient [18]. Although this is considered to be caused by the large difference in mobility between the topological surface and the bulk carriers in 3D-TI, the discussion remains still ambiguous due to the coexisting electronic states of bulk and surface. Considering the situation described so far, accurate discussions on the separate contributions as a function of thickness ($d$) and temperature ($T$) to be viewed simultaneously from the two complementary experimental observation of Hall and Seebeck coefficients are indeed important.

Here, we report our systematic measurements of a set of important electrical transport of sheet resistance ($R_\square$), SdH quantum oscillations, Hall coefficient ($R_H^{2D}$), and Seebeck coefficient ($S$) as a function of both $d$ and $T$ using high quality BSTS single crystal thin films. In order to make unambiguously quantitative discussions on the contribution and the differentiation between the topological surface and the bulk, we grow 3D-TI BSTS thin films with thickness ranging from 5 to 75 nm by employing non-catalytic vapor phase crystal growth reported elsewhere [14,19]. $R_\square$ and $S$ of BSTS films employed in the present experiments show a systematic shift from the bulk/surface coexisting regime to the surface dominant one with a reduction in $d$. $S$ of thinner films clearly shows a linear $T$-dependence with negative polarity from 300 K to 2K, indicating a surface dominant transport of metallic n-type surface carriers in a wide-temperature region. The suppression of the bulk carrier in thin films is quantitatively discussed based on the carrier densities of $n_{Hall}^{2D}$ and $n_{SdH}$ to be evaluated by $R_H^{2D}$ and SdH measurements, respectively. The discrepancy between $n_{Hall}^{2D}$ and $n_{SdH}$ experimentally determined by the two methods becomes smaller and approaches to be constant as $d$ of 3D-TI decreases. The value of $n_{SdH}$ intriguingly approaches to a half value of $n_{Hall}^{2D}$ as $d$ is decreases. We propose that $S$ can be a very sensitive and accurate convenient probe even in the absence of $B$ and at high $T$, and can be employed in place of the combined sophisticated accurate tools of $R_H^{2D}$ and SdH for TIs in order to judge whether the surface dominant electronical transport is realized.

## 2. Experimental

BSTS single crystal thin films with 1 cm² large in size were grown on mica substrate with a catalyst-free epitaxial physical vapor deposition (PVD) method using a dual-quartz tube system, the details of which were reported elsewhere [14,19]. First, a highly insulating $Bi_{1.5}Sb_{0.5}Te_{1.7}Se_{1.3}$ single crystal was synthesized as a source material. The purity of the elements employed for single crystal growth was Bi (5N), Sb (5N), Te (5N), and Se (5N). The source material was then set into a dual-quartz tube system, and the system was evacuated at $10^{-1}$ Pa with a vacuum pump. A mica substrate was located at the other end of the dual quartz tube to grow BSTS single crystal thin films with various thicknesses. The quality of the grown films was characterized by energy dispersive X-ray (EDX) spectroscopy, Raman spectroscopy and X-Ray diffraction (XRD). The thickness of the film was measured by atomic-force-microscopy (AFM).

The resistivity and hall measurements were carried out by a common five probe method using the Physical Properties Measurement System (PPMS, Quantum Design). A magnetic field of 0 to ±9 T perpendicular to the film surface was applied for Hall and magnetoresistance measurements. For the measurement of Seebeck coefficient, a home-built device was used as described elsewhere [17].

## 3. Results

### 3.1 Electrical resistivity.

Figure 1(a) shows $T$ evolution of the 2D sheet resistances ($R_\square$) of five BSTS thin films with different thicknesses (75 nm, 36 nm, 14 nm, 7 nm, and 5 nm). The observed values of $R_\square$ at 300 K are $R_\square$=1.6 (75 nm), 4.3 (36 nm), 14.5 (14 nm), 12.9 (7 nm) and 14.8 kΩ (5 nm). These high $R_\square$ values can ensure that good bulk insulation is realized in our BSTS thin films. It is important to see that $R_\square$ at 300 K shows a large increase in value with a decrement in film thickness from 75 nm to 14 nm, while no significant difference was observed below 14 nm. For the thick film of 75 nm, a typical $T$ dependence similar to that of insulating bulk specimen was observed, where $R_\square$ reached the maximum at around 105 K and started to decrease as $T$ became further low. The insulating property gradually smeared out for the 36 nm film, and by reducing $d$ to be less, an intrinsic metallic $T$ dependence of the nontrivial metallic TSDS emerged over the entire $T$ range.

Figure 1(b) shows the thickness dependence of sheet conductance ($G_\square$) of BSTS films at 300 K and

2 K. In both temperatures, $G_\square$ shows a markedly different behavior above and below the critical thickness of $d_c$=14 nm; a linear increase with an increase in $d$ was obseved above $d_c$, while it became nearly constant below $d_c$. The linear $d$ dependent term in the equation can be considered to be the contribution of the bulk carriers and the constant term can be ascribed to that of the surface carriers. Employing a two-layer parallel connection circuit model as $G_\square = G_\square^S + \sigma_b d$, where $G_\square^S$ and $\sigma_b$ are the sheet conductance of the topological surface and the bulk conductivity, the bulk conductivity was estimated to be $\sigma_b = 52.4$ S cm$^{-1}$ at 2 K and 92.5 S cm$^{-1}$ at 300 K. The decrease in value with temperature can be ascribed to the suppression due to the thermally activated carriers existing in the bulk. On the other hand, $G_\square^S$ below 14 nm increases monotonically from 0.71 (300 K) to $1.02 \times 10^{-4}$ S (2 K), which value at 2 K is comparable with previous reports [10,14,15], suggesting a typical behavior observed in a metal. A similar constant sheet conductance was observed in the entire temperature range of 300 K to 2 K, indicating an experimental fact that a topological layer with two-dimensional Dirac carriers exists with ineligible dependence on $d$ from the viewpoint of electrical transport.

## 3.2 Seebeck coefficient

Figure 2 shows $T$ dependence of Seebeck coefficient ($S$) for five BSTS thin films. The 75 nm-BSTS showed a p-type $S$ with a nonlinear $T$ dependence in positive charge polarity with a maximum value of 193 μVK$^{-1}$ at 300 K, being followed by a sign change to a negative $S$ at 100 K. This experimental result indicates that the dominant carriers evidently change from holes to electrons by decreasing $T$ in the case of the 75 nm-BSTS, being consistent with the observations of $R_\square$ as described earlier. The nonlinear $T$ dependence of $S$ can frequently be observed in semiconductors when carriers are generated in the bulk band via thermal excitations to the upper impurity trapping levels. By reducing $d$ to 36 nm, the value of $S$ decreases to 44 μVK$^{-1}$ and the sign of $S$ changed to be negative even at a higher $T$ of 193 K. On further reduction in $d$, the 14, 7, and 5 nm-BSTSs showed a nearly $d$-independent negative S in the entire $T$ region with linear $T$ dependence, importantly indicating a fact that electrons, but not holes, are dominant in metallic TSDS [17]. The values of $S$ for BSTS films below 14 nm were almost independent of $d$, which is consistent with the results of $R_\square$ in Fig. 1b as described earlier. Judging from the polarity and the linearity of $S$, the surface dominant transport in a wide $T$ range from 300 K to 2 K can be realized around at $d_c$=14 nm.

## 3.3 Hall coefficient

In order to confirm the surface dominant transport quantitatively, we carried out a combined-measurement of Hall resistivity and SdH oscillations. Figure 3(a) and (b) show the magnetic field ($B$) dependences of transverse sheet resistance ($R_{yx}$) of BSTS films at 300 K and 2 K, respectively. In both temperatures, $R_{yx}$ shows a negative slope for all film thicknesses with almost a linear $B$ dependence. The two-dimensional hall coefficient ($R_H^{2D}$) was estimated from a linear fitting under low $B$ from -1 T to 1 T for several temperatures as shown in Fig. 3(c). It is noticed that the sign of $R_H^{2D}$ was always negative, indicating the dominance of electrons of TSDS as a transport carrier.

The 2D-carrier density ($n_{Hall}^{2D}$) was estimated from $R_H^{2D}$ as shown in Fig. 3(d). For the 75 nm-BSTS, $n_{Hall}^{2D}$ shows a strong $T$ dependence, where $n_{Hall}^{2D}$ decreases exponentially from $2.5 \times 10^{14}$ cm$^{-2}$ at 300 K to $3.1 \times 10^{13}$ cm$^{-2}$ at 2 K. The $T$ dependence of $n_{Hall}^{2D}$ became much weaker by reducing the film thickness and reached a nearly constant value of $6.3 \times 10^{12}$ cm$^{-2}$ at 5 nm in thickness. The strong $T$ dependence of $n_{Hall}^{2D}$ could be due to the influence of the bulk carriers, which reduces the intrinsic Hall coefficient ($R_H^{2D}$) of the surface channels and consequently leads to overestimation of $n_{Hall}^{2D}$ at high $T$s. Nearly $T$-independent $n_{Hall}^{2D}$ for both the 5 and 7 nm-BSTSs indicates that the contribution of bulk carriers to the electrical transport becomes negligibly small in these thicknesses and this conclusion is consistent with $R_\square$ and $S$ measurements as described earlier.

## 3.4 Quantum oscillation

SdH quantum oscillations of 75 nm and 5 nm-BSTS observed at 2 K are shown in Fig. 4. Figure 4(a) and (c) were the $\Delta R$-1/$B$ plots obtained by correcting the background with polynomial fitting, where clear SdH oscillations as a function of 1/$B$ can be seen. These quantum oscillations were observed for all BSTS films. By carrying out fast Fourier transformation (FFT) of $\Delta R$-1/$B$ plots, two specific components of $B_F$ =32.4 and 56.8 were revealed for the 75 nm-BSTS to be compared with one component of $B_F$=123.9 for the 5 nm-BSTS as shown in Fig. 4(b) and (d).

For more clear understanding, Fan-diagram plots were made for 75 and 5 nm-BSTSs as shown in Fig.4 (e). The experimental line in black indicates the peak and valley positions for the 75 nm BSTS, and the red one is for the 5 nm-BSTS. Both black and red lines were obtained from the linear fitting of each plot using a curvature value of 1/$B_F$ evaluated from FFT analyses. Two conductive channals for the 75 nm film have different berry phases of $\beta = 0.00$ ($B_F$=56.8) and 0.55 ($B_F$=32.4), and the value of conductive channel of 5 nm film is $\beta = 0.64$ ($B_F$=123.9). Importantly, carriers of both non-trivial TSDS ($\beta$=1/2) and trivial bulk state ($\beta$=0) were observed in the case of the thick 75 nm-BSTS, while only TSDS state was observed in the thin 5 nm-BSTS. The disappearance of the bulk states for the thin 5 nm-BSTS would be reasonable considering the larger reduction of the bulk contribution as $d$ reduces.

## 4. Discussion
### 4.1 Comparison of carrier density of Hall and SdH measurements

Based on the experimental data and their analyses of SdH quantum oscillations descried earlier, the 2D-carrier densities of TSDS were estimated as $n_{SdH} = 7.8 \times 10^{11}$ cm$^{-2}$ for the 75 nm-BSTS and $3.00 \times 10^{12}$ cm$^{-2}$ for the 5 nm-BSTS. It is notice that the carrier density of the 75 nm-BSTS is by two orders in magnitude smaller than that evaluated from Hall measurements ($n_{Hall}^{2D} = 3.1 \times 10^{13}$ cm$^{-2}$), while being in strong contrast the two values were within the same order in the case of the 5 nm-BSTS ($n_{Hall}^{2D} = 5.73 \times 10^{12}$ cm$^{-2}$). These experimental observations provide us the following information. Generally, when a material has two types of conducting channels for n- and p-carriers, linear fitting of the $R_{yx}$ will underestimate the Hall coefficient to lead to an overestimate of the carrier density due to the compensation in sign between electrons and holes. On the other hand, SdH oscillations can extract the intrinsic carrier density for each channel separately, which can accurately be possible by deconvoluting the experimental data corresponding to each channel.

Figure 5 shows the thickness dependence of estimated 2D-carrier densities of TSDS evaluated from Hall measurements ($n_{Hall}^{2D}$) and SdH measurements ($n_{SdH}$) at 2 K. It is clearly seen that the large discrepancy between the two values becomes smaller as the film thickness is reduced. Importantly, the ratio of two carrier densities $n_{Hall}^{2D}/n_{SdH}$ become constant to be two below the film thickness of 7 nm, as shown in the inset of Fig. 5. In 3D-TIs, two conductive surface channels can be considered on the top and the bottom surfaces of TSDS. Therefore, the carrier density revealed by Hall measurements should be the sum of these two surfaces, and its value of twice the $n_{SdH}$ could be reasonable. The contribution of the bulk carriers completely diminishes at the film thickness below 7 nm and intrinsic nontrivial pure m-TSDS can be observed.

### 4.2 The inconsistency of Seebeck and Hall coefficients

It is importantly noted that the sign of $R_H^{2D}$ was always negative even in the thick films of 75 and 36 nm-BSTS, which contradicts to the results obtained from $S$ measurements, where a positive $S$ value was observed for thick films as described earlier. In general, the polarity of Seebeck and Hall coefficients should be consistent, i.e. the positive Seebeck/Hall coefficient corresponds to p-type hole carriers and negative coefficient to n-type electron carriers. The different contribution observed between Hall and

Seebeck can be interpreted in terms of the large difference in carrier mobility between the trivial bulk and the non-trivial surface states of TIs.

Applying a two-parallel circuit model of a bulk and a surface, Seebeck and Hall coefficients can be described as

$$S = \frac{\sigma_b t S_b - G_\square^S S_s}{\sigma_b t + G_\square^S} = -\frac{G_\square^S S_s}{G_\square}\left(1 - \frac{\sigma_b t}{G_\square^S}\frac{S_b}{S_s}\right) \quad (1)$$

$$R_H^{2D} = \frac{\sigma_b t \mu_b - G_\square^S \mu_s}{\left(\sigma_b t + G_\square^S\right)^2} = -\frac{G_\square^S \mu_s}{G_\square^2}\left(1 - \frac{\sigma_b t}{G_\square^S}\frac{\mu_b}{\mu_s}\right) \quad (2)$$

, where $\sigma_b$, $\mu_b$, $S_b$ are the electrical conductivity, the mobility, and the Seebeck coefficient of bulk, $G_\square^S$, $\mu_s$, $S_s$ are the sheet conductance, the mobility, and the Seebeck coefficient of surface, and $t$ is the thickness of films. By compering these two equations, it is clear that the difference in sign between the two coefficients comes from the ratio of $\frac{S_b}{S_s}$ and $\frac{\mu_b}{\mu_s}$. According to the linear fitting analyses of sheet conductance in Fig. 1(b) as described earlier, the ratio of $\frac{\sigma_b t}{G_\square^S}$ is 10 and 5 for 75 and 36 nm-BSTS at 300 K. Using the $S$ value of 193 µVK$^{-1}$ for the 75 nm-BSTS as $S_b$, and -45 µVK$^{-1}$ for the 5 nm-BSTS as $S_s$, the multiplied value becomes $\frac{\sigma_b t}{G_\square^S}\frac{S_b}{S_s}$ = 43 and 21 for the 75 nm and the 36 nm-BSTS, respectively. When this multiplied value is larger than 1, a positive $S$ can be observed.

On the other hand, Dirac electrons of surface have a large mobility due to the prohibition of backward scattering, resulting in a greatly smaller value of $\frac{\mu_b}{\mu_s}$. The mobility of TSDS can be estimated from the SdH oscillations using the following equation:

$$\Delta R_{xx} = A \exp(-\pi/\mu^* B) \cos[2\pi(B_F/B + 1/2 + \beta)] \quad (3)$$

where $A$ is the amplitude, $\mu^*$ is the carrier mobility, $B_F$ is the periodic frequency of the oscillations, and $\beta$ is the Berry phase [10,20,21]. Figure 6 shows the fitting result for the 7 nm-BSTS as a typical example, where fitting was carried out by employing the evaluated $B_F$ and $\beta$ value from FFT and fan-diagram plot analyses ($B_F$ = 77.5 and $\beta$ = 0.35). The value of $\mu^*$ for TSDS in the Dirac electron pocket was evaluated to be 1078 cm$^2$ V$^{-1}$ s$^{-1}$, which is similar to the previous measurements [14]. A typical value of the bulk mobility of BSTS is only several tens cm$^2$ V$^{-1}$ s$^{-1}$ [15, 22-24], resulting in $\frac{\mu_b}{\mu_s}$ ~ 0.01 and a negative value of $R_H^{2D}$. The minus value of $R_H^{2D}$ in 300 K indicates that the mobility of surface electron carriers is still quite larger than that of the bulk holes even in the high $T$ region.

## 5. Conclusion

We systematically observed a whole set of electrical transport: longitudinal resistivity ($R$), transverse Hall ($R_H^{2D}$) coefficient, Seebeck ($S$) coefficient, and SdH quantum oscillations, for high quality 3D-TI BSTS as a function of temperature ($T$) and film thickness ($d$) in order to clarify the contributions of both bulk and TSDS carriers. Accurate quantitative discussions were successfully made on the carrier densities of $n_{SdH}$ and $n_{Hall}^{2D}$ estimated by both measurements $R_H^{2D}$ and SdH oscillations. A discrepancy which has been a debate so far among researchers was seen between the two types of measurements. The $d$ dependences of sheet conductance ($R_\square$) and $S$ coefficient clearly show that the semiconducting hole carriers stemming from the bulk states can reasonably be suppressed with a reducing in $d$, and a topological surface dominant transport can be obtained. While $R_\square$ and S coefficient were apparently contributed from both the bulk and the surface in the case of thick BSTS films, $R_H^{2D}$ differently showed single type carriers arising only from the TSDS, even when the contribution from the bulk carriers cannot be negligible. The situation was interpreted in terms of the lager difference in mobility between the surface and the bulk. Cautiously, the carrier density ($n$) of TSDS estimated by $R_H^{2D}$ provides an overestimation due to the additional influence of the bulk carriers. According to the accurate quantitative comparison of $n$ between Hall coefficient ($n_{Hall}^{2D}$) and SdH oscillations ($n_{SdH}$), we clarified that such overestimate gradually becomes small as $d$ decreases and intriguingly $n_{Hall}^{2D}$ approaches to a value of almost the twice of $n_{SdH}$. We propose that Seebeck coefficient can be a very powerful and useful probe in order to estimate quantitatively the carrier number and polarity intrinsically contributing to the electrical transport of TIs without employing $B$ and low $T$.

**Acknowledgments**

This work was supported in part by a Grant-in-Aid for Scientific Research from the Ministry of Education, Culture, Sports, Science and Technology (MEXT), JSPS KAKENHI Grants No. 17K14329, 18H04471, 17-18H05326, 18H04304, 18H03883, and 18H03858) and thermal management of CREST, JST. This work was sponsored by research grants from Yazaki Memorial Foundation and The Iwatani Naoji Foundation's Research Grant. S.Y.M. thanks Tohoku University Interdepartmental Doctoral Degree Program for Multi-dimensional Materials Science Leaders for financial support. The research is partly carried out by the support of the World Premier International Research Center Initiative (WPI) from MEXT.

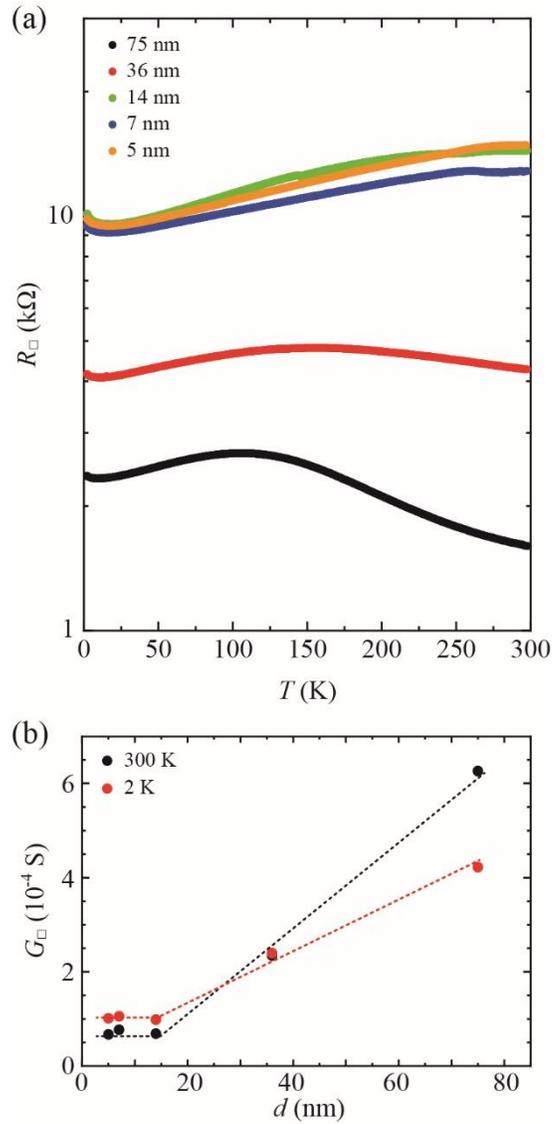

**Figure 1.** Electrical transport properties of BSTS thin films. (a) Temperature dependence of sheet resistance of BSTS thin films of 75, 36, 14, 7, and 5 nm. (b) Film thickness dependence of total conductance at 300 K (black circles) and 2 K (red circles).

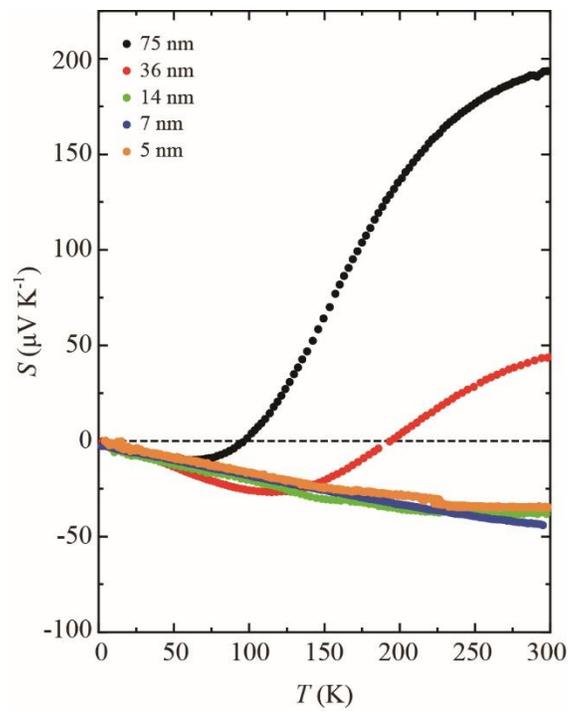

**Figure 2.** Seebeck coefficient of BSTS thin films. Temperature dependence of Seebeck coefficient (*S*) of BSTS thin films of 75, 36, 14, 7, and 5 nm.

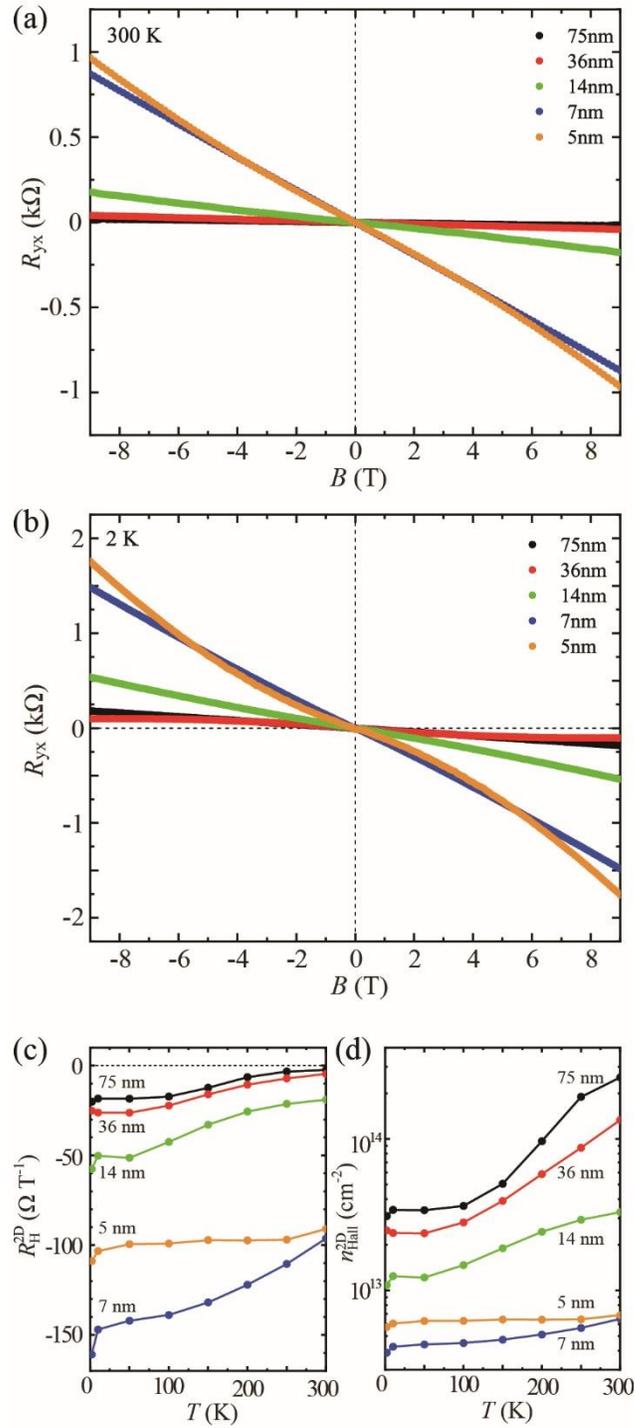

**Figure 3.** Thickness dependence of Hall resistance of BSTS thin films. Temperature dependence of Hall resistance ($R_{yx}$) of BSTS thin films of 75, 36, 14, 7, and 5 nm measured at 300 K for (a) and 2 K for (b). (c) two-dimensional Hall coefficient of each film estimated by linear curve fitting within the range of -1 T to 1 T. (d) two-dimensional carrier density of each film.

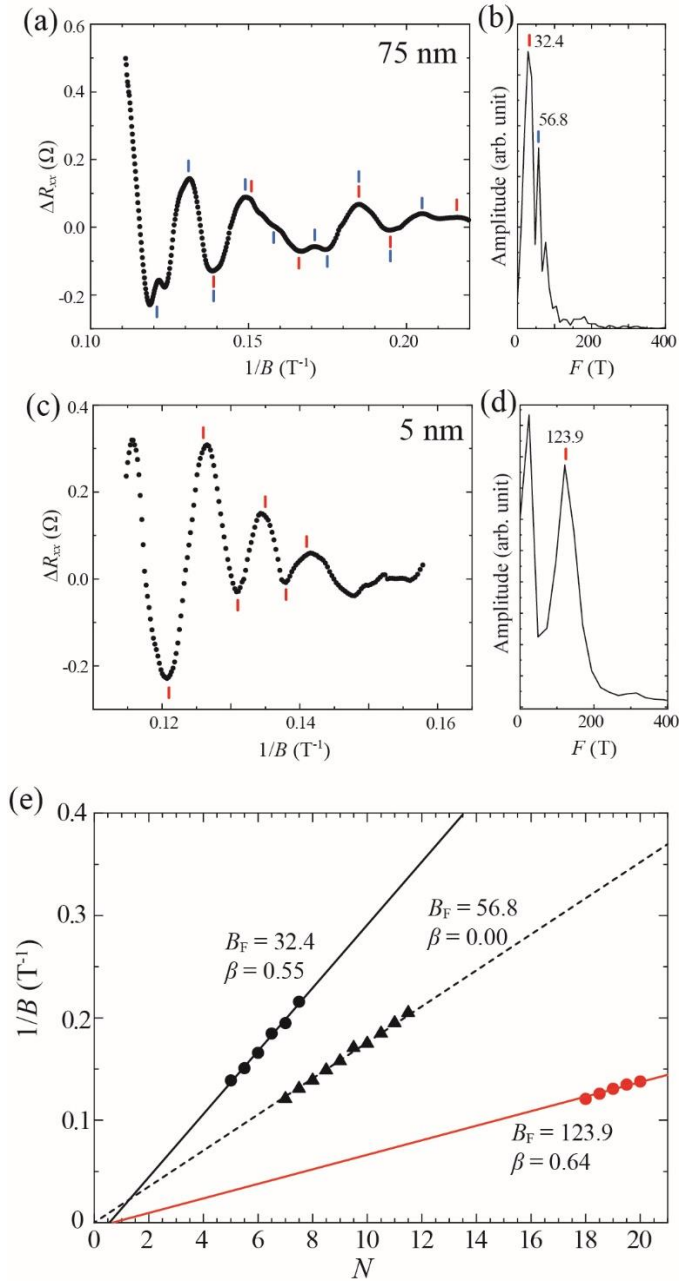

**Figure 4.** Shubnikov-de-Haase oscillations of BSTS thin films. (a), (c) SdH oscillations of 75 and 5 nm-BSTS films, respectively. Black and red bars indicate the peak and the valley positions. (b), (d) Fast Fourie Transfer (FFT) of the SdH oscillations of 75 and 5 nm BSTS, respectively. (e) Fan-diagram plots of 75 and 5 nm BSTS films, respectively. Black circles and triangles represent the peak and valley positions of 75 nm-BSTS in (a), where circles (triangles) corresponds to the non-trivial TSDS (trivial bulk state) in 3D-TI. Red circles are the peak and valley positions of 5 nm-BSTS in (c) correspond to non-trivial TSDS. The bold and dotted lines are the linear fitting curves for each electronic state.

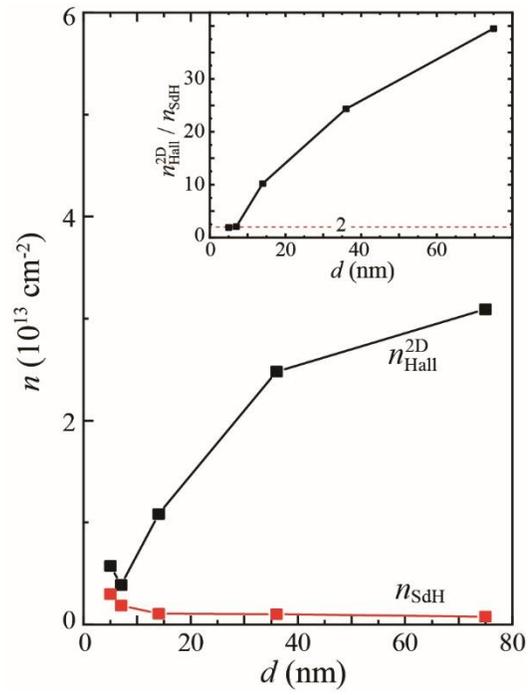

**Figure 5.** Comparison of carrier density estimated from Hall and SdH oscillation measurements. The thickness dependence of two-dimensional carrier density at 2 K estimated by Hall coefficient (Black) and SdH oscillation (Red). The inset represents the thickness dependence of the ratio $n_{SdH}/n_{Hall}^{2D}$.

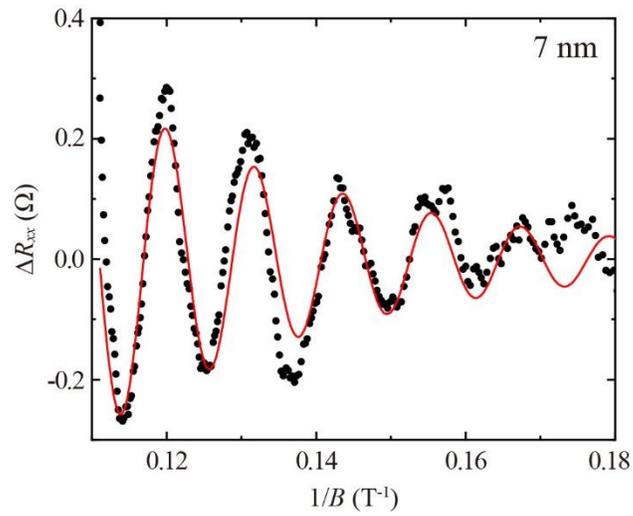

**Figure 6.** Fitting analysis of Shubnikov-de-Haase oscillations of 7 nm-BSTS thin film. An example of fitting analysis of SdH oscillation of 7 nm-BSTS film. Red curve represents the result of fitting analysis using equation (3) in the text.